\documentclass[showpacs,preprintnumbers,10pt,twocolumn]{revtex4}%
\usepackage{epsfig,float}
\usepackage{graphicx}
\usepackage{amssymb}
\usepackage{amsfonts}
\usepackage{amsmath}
\usepackage{dcolumn}
\usepackage{bm}
\usepackage{revsymb}%
\providecommand{\U}[1]{\protect\rule{.1in}{.1in}}
\begin{document}
\title{Reply to \textquotedblleft Comment on `Ratchet universality in the presence of
thermal noise'\ \textquotedblright}
\author{Pedro J. Mart\'{\i}nez$^{1}$ and Ricardo Chac\'{o}n$^{2}$}
\affiliation{$^{1}$Departamento de F\'{\i}sica Aplicada, E.I.N.A., Universidad de Zaragoza,
E-50018 Zaragoza and Instituto de Ciencia de Materiales de Arag\'{o}n,
CSIC-Universidad de Zaragoza, E-50009 Zaragoza, Spain, EU }
\affiliation{$^{2}$Departamento de F\'{\i}sica Aplicada, Escuela de Ingenier\'{\i}as
Industriales, Universidad de Extremadura, Apartado Postal 382, E-06006
Badajoz, Spain, EU}
\date{\today}

\begin{abstract}
The Comment by Quintero \textit{et al}.~does not dispute the central result of
our paper [Phys. Rev. E \textbf{87}, 062114 (2013)] which is a theory
explaining the interplay between thermal noise and symmetry breaking in the
ratchet transport of a Brownian particle moving on a periodic substrate
subjected to a temporal biharmonic excitation $\gamma\left[  \eta\sin\left(
\omega t\right)  +\alpha\left(  1-\eta\right)  \sin\left(  2\omega
t+\varphi\right)  \right]  $. In the Comment, the authors claim, on the sole
basis of their numerical simulations for the \textit{particular }case
$\alpha=2$, that \textquotedblleft there is no such universal force waveform
and that the evidence obtained by the authors otherwise is due to their
particular choice of parameters.\textquotedblright\ Here we demonstrate by
means of theoretical arguments and additional numerical simulations that all
the conclusions of our original article are preserved.

\end{abstract}

\maketitle

The foregoing Comment by Quintero \textit{et al}.~[1] offers some criticisms
on certain particular (numerical) aspects of our previous paper [2], in which
we studied theoretically and numerically a universal model -- a Brownian
particle moving on a periodic substrate subjected to a biharmonic excitation,%
\begin{align}
\overset{.}{x}+\sin x  &  =\sqrt{\sigma}\xi\left(  t\right)  +\gamma
F_{bihar}\left(  t\right)  ,\nonumber\\
F_{bihar}\left(  t\right)   &  \equiv\eta\sin\left(  \omega t\right)
+\alpha\left(  1-\eta\right)  \sin\left(  2\omega t+\varphi\right)  , \tag{1}%
\end{align}
where $\gamma$ is an amplitude factor, and the parameters $\left(  \eta
\in\left[  0,1\right]  ,\alpha>0\right)  $ and $\varphi$ account for the
relative amplitude and initial phase difference of the two harmonics,
respectively, while $\xi\left(  t\right)  $ is a Gaussian white noise with
zero mean and $\left\langle \xi\left(  t\right)  \xi\left(  t+s\right)
\right\rangle =\delta\left(  s\right)  $, and $\sigma=2k_{b}T$ with $k_{b}$
and $T$ being the Boltzmann constant and temperature, respectively.

Quintero \textit{et al}.~state in their Comment's abstract that
\textquotedblleft The authors claim that their simulations prove the existence
of a universal waveform of the external force which optimally enhances
directed transport, hence confirming the validity of a previous conjecture put
forward by one of them in the limit of vanishing noise
intensity.\textquotedblright\ We disagree with this statement. Note that the
existence of such a universal waveform was clearly conjectured for the first
time in Ref.~[3] in the context of a criticality scenario: Optimal enhancement
of directed ratchet transport (DRT) is achieved when maximally effective
(i.e., critical) symmetry breaking occurs. The mathematical proof of the
ratchet conjecture was subsequently completed in Ref.~[4] where such a
universal waveform is shown to be unique for both temporal and spatial
biharmonic forces. This universal waveform is a direct consequence of the
degree of symmetry breaking (DSB) mechanism: It is possible to consider a
quantitative measure of the DSB on which the strength of directed transport by
symmetry breaking must depend. Such a theory of the ratchet universality has
been applied to predict successfully the behavior of two kinds of soliton
ratchets [5,6]. The theory demonstrated in our original article [2] uses,
among other theoretical ideas, ratchet universality to explain the interplay
between thermal noise and symmetry breaking in the ratchet transport of Eq.
(1). The authors of the Comment neither take into account nor do they cite Refs.~[3,5,6].

All the criticisms made by Quintero \textit{et al}.~are based solely on
numerical simulations of Eq.~(1) for the \textit{particular} case $\alpha=2$.
However, the theoretical discussion of the structural stability of the ratchet
scenario under changes in the parameter $\alpha$ (cf.~the text and Fig.~1(c)
in Ref.~[2]) is \textit{essential} to understand the theory demonstrated in
our original article. Indeed, the authors claim that the numerical results
shown in Fig.~1 of their Comment allow one to reach their two main conclusions:

(a) \textquotedblleft First of all, our results are not compatible with the
existence of an optimal force waveform [...] The apparent (approximate)
confirmation of the prediction $\eta_{opt}=4/5$ seems to arise from the
specific choice of simulation parameters made by the
authors.\textquotedblright\ As anticipated above, this conclusion arises from
a misunderstanding on the part of the authors concerning the theory
demonstrated in our original article [2]. Specifically, we demonstrated in
Ref.~[2] that the effect of thermal noise on the purely deterministic ratchet
scenario can be understood as an effective noise-induced change of the
potential barrier which allows one to understand the existence and behavior of
the deviation $\Delta\eta\left(  \alpha\right)  \equiv\eta_{opt}\left(
\alpha\right)  -\eta_{opt}^{\sigma>0}\left(  \alpha\right)  $ as $\alpha$ is
changed, while keeping constant the remaining parameters. Clearly, as
mentioned in Ref.~[2], the effect of noise on the DRT depends on the amplitude
of the biharmonic excitation while keeping constant the remaining parameters.
This means that the deviation $\Delta\eta\left(  \alpha\right)  $ will depend
in its turn on the amplitude factor $\gamma$, while our theory, which was
shown to be numerically confirmed for the particular value $\gamma=2$ just as
an illustrative example, holds for other values of the amplitude factor.
Indeed, Fig.~1 shows an additional illustrative example for $\gamma=6$. One
finds that the range where the DSB mechanism dominates over the thermal
interwell activation (TIA) mechanism is larger for $\gamma=6$ than for
$\gamma=2$, as expected from our theory.

\begin{figure}
\begin{center}
\epsfig{file=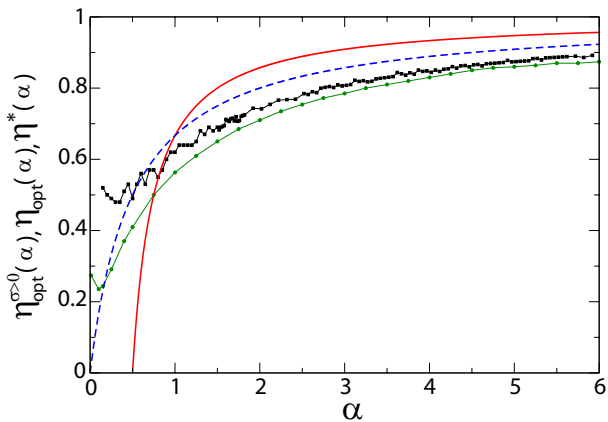,width=0.45\textwidth}
\end{center}
\caption{Fig.~1 (Color online) Value of $\eta$ where the average velocity is maximum,
$\eta_{opt}^{\sigma>0}$, versus $\alpha$ [cf.~Eq.~(1)] for $\varphi
=\varphi_{opt}\equiv\pi/2,\omega=0.08\pi,\sigma=2$, and two values of the
amplitude factor: $\gamma=2$ (squares) and $\gamma=6$ (dots). Also plotted is
the theoretical prediction for the purely deterministic case $\eta
_{opt}\left(  \alpha\right)  \equiv2\alpha/\left(  1+2\alpha\right)  $ (dashed
line) and the function $\eta^{\ast}(\alpha)\equiv\left(  4\alpha-2\right)
/\left(  4\alpha-1\right)  $ (solid line, cf.~Ref.~[2]).
}
\label{figura1}
\end{figure}

(b) \textquotedblleft Secondly, although we cannot decrease $\gamma$ below
$0.8$ without introducing too much uncertainty, the figure clearly illustrates
that the trend of the value of $\eta$ which optimizes $v$ [for the average
velocity] is toward the value $2/3$ which all theories predict in their range
of validity, i.e., \textit{in the limit of weak external forces}%
.\textquotedblright\ Note that this \textit{extrapolation argument}
invalidates itself: Directed ratchet transport is not properly defined in the
limit of weak (with respect to the potential barrier) external forces in the
presence of finite but arbitrary thermal noise, which is indeed confirmed in
Fig.~1 of the Comment by the absence of numerical results for values of
$\gamma\in\left]  0,0.8\right[  $, and \textit{wovon man nicht sprechen kann,
dar\"{u}ber mu}ss \textit{man schweigen }[7].

The rest of the Comment presents a few examples of the application of a
numerical fitting method for the average velocity which is solely subjected to
obvious constraints regarding the breaking of relevant symmetries in Eq.~(1).
Two observations concerning this method are in order. First, as is well known,
numerical and experimental data providing the dependence of the average
velocity on relevant parameters, such as $\eta$ and $\varphi$, in many
ratchets fit generally into soft curves that exhibit few extrema. It is thus
not so surprising that the authors' fitting method yielded good fits having so
many free adjustable parameters (see, for example, Fig.~3(b) in the Comment).
Even just a few parameters are enough to fit an elephant [8]! And second, the
authors' fitting method is not a physical theory since it has no predictive
power at all $-$precisely because of its dependence on many adjustable
parameters. This is in sharp contrast to the theory of the ratchet
universality which predicted and explained, for example, why the directed
soliton current is dependent on the number of atoms in the DRT of bright
solitons formed in a quasi-one-dimensional Bose-Einstein condensate [6], as
well as the strong enhancement of DRT of topological solitons in
Frenkel-Kontorova chains due to the introduction of phase disorder into the
asymmetric periodic driving [5]. Finally, the authors again misunderstand the
ratchet universality when commenting on their results shown in their Fig.~4
(final paragraph): The ratchet universality does not predict a sinusoidal
dependence of the average velocity on the initial phase difference $\varphi$
for general values of the amplitude factor. This is only valid for
sufficiently small amplitudes (cf.~Ref.~[3]). Only the optimal values of
$\varphi$ are predicted to be universal. In the present case, these values are
$\varphi_{opt}=\left\{  \pi/2,3\pi/2\right\}  $ (cf.~Ref.~[3]) which are
indeed confirmed by the numerical results shown in Fig.~4 of the Comment.

P.J.M. acknowledges financial support from the Ministerio de Econom\'{\i}a y
Competitividad (MECC, Spain) through Project No. FIS2011-25167 cofinanced by
FEDER funds. R.C. acknowledges financial support from the MECC, Spain, through
Project No. FIS2012-34902 cofinanced by FEDER funds, and from the Junta de
Extremadura (JEx, Spain) through Project No. GR10045.

\end{document}